\documentstyle[preprint,tighten,aps,amstex,graphicx,floats]{revtex}

\begin{document}

\thispagestyle{empty}

\marginparwidth 1.cm
\setlength{\hoffset}{-1cm}
\newcommand{\mpar}[1]{{\marginpar{\hbadness10000%
                      \sloppy\hfuzz10pt\boldmath\bf\footnotesize#1}}%
                      \typeout{marginpar: #1}\ignorespaces}
\def\mda{\mpar{\hfil$\downarrow$\hfil}\ignorespaces}
\def\mua{\mpar{\hfil$\uparrow$\hfil}\ignorespaces}
\def\mla{\marginpar[\boldmath\hfil$\rightarrow$\hfil]%
                   {\boldmath\hfil$\leftarrow $\hfil}%
                    \typeout{marginpar: $\leftrightarrow$}\ignorespaces}

\def\ba{\begin{eqnarray}}
\def\ea{\end{eqnarray}}
\def\bq{\begin{equation}}
\def\eq{\end{equation}}

\renewcommand{\abstractname}{Abstract}
\renewcommand{\figurename}{Figure}
\renewcommand{\refname}{Bibliography}

\newcommand{\eg}{{\it e.g.}\;}
\newcommand{\ie}{{\it i.e.}\;}
\newcommand{\etal}{{\it et al.}\;}
\newcommand{\ibid}{{\it ibid.}\;}

\newcommand{\mx}{M_{\rm SUSY}}
\newcommand{\pt}{p_{\rm T}}
\newcommand{\et}{E_{\rm T}}
\newcommand{\del}{\varepsilon}
\newcommand{\sla}[1]{/\!\!\!#1}
\newcommand{\fb}{{\rm fb}}
\newcommand{\gev}{{\rm GeV}}
\newcommand{\tev}{{\rm TeV}}
\newcommand{\abi}{\;{\rm ab}^{-1}}
\newcommand{\fbi}{\;{\rm fb}^{-1}}

\newcommand{\zpc}[3]{${\rm Z. Phys.}$ {\bf C#1} (#2) #3}
\newcommand{\epc}[3]{${\rm Eur. Phys. J.}$ {\bf C#1} (#2) #3}
\newcommand{\npb}[3]{${\rm Nucl. Phys.}$ {\bf B#1} (#2)~#3}
\newcommand{\plb}[3]{${\rm Phys. Lett.}$ {\bf B#1} (#2) #3}
\renewcommand{\prd}[3]{${\rm Phys. Rev.}$ {\bf D#1} (#2) #3}
\renewcommand{\prl}[3]{${\rm Phys. Rev. Lett.}$ {\bf #1} (#2) #3}
\newcommand{\prep}[3]{${\rm Phys. Rep.}$ {\bf #1} (#2) #3}
\newcommand{\fp}[3]{${\rm Fortschr. Phys.}$ {\bf #1} (#2) #3}
\newcommand{\nc}[3]{${\rm Nuovo Cimento}$ {\bf #1} (#2) #3}
\newcommand{\ijmp}[3]{${\rm Int. J. Mod. Phys.}$ {\bf #1} (#2) #3}
\renewcommand{\jcp}[3]{${\rm J. Comp. Phys.}$ {\bf #1} (#2) #3}
\newcommand{\ptp}[3]{${\rm Prog. Theo. Phys.}$ {\bf #1} (#2) #3}
\newcommand{\sjnp}[3]{${\rm Sov. J. Nucl. Phys.}$ {\bf #1} (#2) #3}
\newcommand{\cpc}[3]{${\rm Comp. Phys. Commun.}$ {\bf #1} (#2) #3}
\newcommand{\mpl}[3]{${\rm Mod. Phys. Lett.}$ {\bf #1} (#2) #3}
\newcommand{\cmp}[3]{${\rm Commun. Math. Phys.}$ {\bf #1} (#2) #3}
\newcommand{\jmp}[3]{${\rm J. Math. Phys.}$ {\bf #1} (#2) #3}
\newcommand{\nim}[3]{${\rm Nucl. Instr. Meth.}$ {\bf #1} (#2) #3}
\newcommand{\prev}[3]{${\rm Phys. Rev.}$ {\bf #1} (#2) #3}
\newcommand{\el}[3]{${\rm Europhysics Letters}$ {\bf #1} (#2) #3}
\renewcommand{\ap}[3]{${\rm Ann. of~Phys.}$ {\bf #1} (#2) #3}
\newcommand{\jhep}[3]{${\rm JHEP}$ {\bf #1} (#2) #3}
\newcommand{\jetp}[3]{${\rm JETP}$ {\bf #1} (#2) #3}
\newcommand{\jetpl}[3]{${\rm JETP Lett.}$ {\bf #1} (#2) #3}
\newcommand{\acpp}[3]{${\rm Acta Physica Polonica}$ {\bf #1} (#2) #3}
\newcommand{\science}[3]{${\rm Science}$ {\bf #1} (#2) #3}
\newcommand{\vj}[4]{${\rm #1~}$ {\bf #2} (#3) #4}
\newcommand{\ej}[3]{${\bf #1}$ (#2) #3}
\newcommand{\vjs}[2]{${\rm #1~}$ {\bf #2}}
\newcommand{\hep}[1]{${\tt hep\!-\!ph/}$ {#1}}
\newcommand{\hex}[1]{${\tt hep\!-\!ex/}$ {#1}}
\newcommand{\desy}[1]{${\rm DESY-}${#1}}
\newcommand{\cern}[2]{${\rm CERN-TH}${#1}/{#2}}

\preprint{
\font\fortssbx=cmssbx10 scaled \magstep2
\hbox to \hsize{\hskip.5in 
\raise.1in\hbox{\fortssbx Fermi National Accelerator Laboratory}
\hfill\vtop{\hbox{\bf FERMILAB-Pub-01/295-T}
            \hbox{hep-ph/0110218} } }
}

\title{ 
Top quark associated production of topcolor pions at hadron colliders
} 

\author{
Adam K.\ Leibovich and David Rainwater
} 

\address{ 
Theory Dept., Fermi National Accelerator Laboratory, Batavia, IL, USA
} 

\maketitle 

\begin{abstract}
We investigate the associated production of a neutral physical pion with 
top quarks in the context of topcolor assisted technicolor. We find that 
single-top associated production does not yield viable rates at either 
the Tevatron or LHC. $t\bar{t}$-associated production at the Tevatron is 
suppressed relative to Standard Model $t\bar{t}H$, but at the LHC is 
strongly enhanced and would allow for easy observation of the main decay 
channels to bottom quarks, and possible observation of the decay to gluons.
\end{abstract} 

\vspace{5mm}



\section{Introduction}\label{sec:intro}

Hadron colliders are machines extremely well-suited to study the forefront 
problem of electroweak symmetry breaking (EWSB) and fermion mass generation. 
Fermilab's Tevatron, now engaged in Run II, has significant potential to 
discover a light Standard Model (SM) or Minimal Supersymmetric Standard Model 
(MSSM) Higgs boson, with mass up to about $M_H \lesssim 130$~GeV~\cite{H-Tev2}. 
However, it will have very little capability to determine the overarching model 
that governs EWSB if a Higgs candidate is observed. 
The CERN Large Hadron Collider (LHC), on the other hand, will have considerably 
expanded capability to discover and measure almost all the quantum properties 
of a SM Higgs of any mass or several of the MSSM Higgs bosons over the entire 
MSSM parameter space~\cite{H-LHC,H-prop,LHC-high-pT,gunion,HVVstruct}. 
While this is certainly very promising for future studies of EWSB, very little 
attention has been given recently to non-SM/MSSM theories of gauge boson and 
fermion mass generation. 

Of particular concern to us are the more modern dynamical models of EWSB.  
While dynamical models have historically had many theoretical problems 
as well as conflicts with data, and broad classes have been ruled out, 
there are still viable models worthy of investigation in light of the 
capabilities of the current generation of experiments.  We address here 
the theory of topcolor assisted technicolor (TC2)~\cite{TC2}, 
specifically type I~\cite{BBHK}.  This model is still consistent with
experiment~\cite{hill}.  We first outline the model in 
Sec.~\ref{sec:model}, discuss the phenomenology of the model in 
Sec.~\ref{sec:pheno}, and then present conclusions and the outlook for 
upcoming experiments.  Details of some of the analytical calculations 
are presented in the Appendices.


\section{The topcolor assisted technicolor model}\label{sec:model}

Dynamical theories of fermion mass generation, the most viable of which 
is extended technicolor (ETC), typically have difficulty accommodating 
the large top quark mass. 
TC2 was proposed to assuage this problem, by having two separate 
strongly interacting sectors. One (topcolor, or TC) provides for the 
large top quark mass but has comparatively little contribution to EWSB, 
while the other (ETC) is responsible for the bulk of EWSB but contributes 
almost nothing to $m_t$.  Details of TC2 may be found in Ref.~\cite{TC2}. 
Here, we briefly review the characteristics most relevant for discussion 
of its phenomenology. 

Topcolor gauge interactions cause top quark pair condensation at some scale 
$\Lambda$ via a strong four-fermion interaction
\begin{equation}
{g^2\over\Lambda^2} \bar{\psi}_L t_R \bar{t}_R \psi_L,
\label{eq:4f}
\end{equation}
where the fields are the SM third generation matter fields in $SU(2)$
doublet and singlet representation.  The resulting chiral symmetry
breaking yields a set of Goldstone bosons.  The interaction of these
bosons and the condensate may be written as an $SU(2)$ field $\Phi$ in
exponential form:
\begin{equation}
\Phi_{TC} = e^{i \vec\tau\cdot\vec\pi / f_\pi} 
\binom{(f_\pi + H_{TC})/\sqrt{2}}{0},
\label{eq:primal}
\end{equation}
where $f_\pi$ is the vacuum expectation value (vev) of the top quark pair 
condensate, and $\vec\tau$ are the Pauli matrices.
Note that the hypercharge of $\Phi$ is $-1$. 
Type I topcolor contains an extra $U(1)$ which tilts the fermion interaction 
to disallow condensation of a $b\bar{b}$ condensate as well. 
A similar condensation $<\bar{T}_L T_R>$ of technifermions occurs in the ETC 
sector, with its own vev $v_T$, and one may write the $SU(2)$ doublet 
$\Phi_{ETC}$ in the same form.

The Pagels-Stokar formula~\cite{pagels} gives the value of the vev $f_\pi$ in 
terms of the number of topcolors, the top quark mass, and the scale at which 
the condensation occurs:
\begin{equation}
f^2_\pi \, \simeq \, 
{N_c\over 16\pi^2} \, m^2_t \; 
\biggl[ {\rm ln} \biggl( {\Lambda^2\over m^2_t} \biggr) + K \biggr],
\label{eq:pagels}
\end{equation}
where $K$ is a constant of order 1. 
For condensation around the EWSB scale of 1~TeV, $f_\pi \simeq 60$~GeV, but it 
should be understood that this is only a rough guide, and $f_\pi$ may in fact 
be somewhat lower or higher, say in the range $40-80$~GeV. 
Allowing $f_\pi$ to vary over this range does not qualitatively change our 
conclusions and has only minimal impact on our quantitative results. 
Therefore, we use the value $f_\pi = 60$~Gev throughout our analysis as a 
convenient baseline. 

We linearize the theory and rearrange the pions in two orthogonal linear 
combinations to form the longitudinal degrees of freedom of the weak gauge 
bosons and a triplet of ``top-pions'', $\Pi^{0,\pm}$, which become physical 
degrees of freedom.  (See Appendix~\ref{app:model} for details.) 
The top-pions are analogous to the neutral CP-odd and charged Higgs scalars of 
a two-Higgs doublet model (2HDM), of which the MSSM Higgs sector is a subset. 
Contributions to the top quark mass can come from both sectors, but the model 
assumes that the dominant contribution is from TC. 
The top quark Yukawa term in the Lagrangian, ignoring mixing between the 
two Higgs modes, is written as
\begin{eqnarray}
{\cal L}_{Yuk,t} \; &
= & \; -{1\over\sqrt{2}} 
     \, \biggl( Y_t \, f_\pi + \epsilon_t \, v_T \biggr) \bar{t} t \nonumber\\
& & \; -{1\over\sqrt{2}} 
       \, \biggl( Y_t \, H_{TC} + \epsilon_t \, H_{ETC} \biggr) \bar{t} t 
\nonumber \\
& & \; -{i\over v\sqrt{2}} \biggl( Y_t \, v_T - \epsilon_t f_\pi \biggr) 
                           \, \Pi^0 \bar{t} \gamma^5 t \; .
\label{eq:Yuk-t}
\end{eqnarray}
where $Y_t$ is the TC Yukawa coupling, and $\epsilon_t$ is a small ETC 
contribution. Once $f_\pi$ is fixed, $v_T$ is uniquely determined by the 
EWSB requirement that $f^2_\pi + v^2_T = v^2 \simeq (246 \; {\rm GeV})^2$. 
For $f_\pi = 60$~GeV, we must have $v_T = 239$~GeV.  The measured top mass 
then fixes $Y_t$ to be of order 3-4 for small $\epsilon_t$.  The maximal 
value of $Y_t$ is $Y_{t,max} = 4.1$ occurs when $\epsilon_t = 0$. 
We neglect the effects of flavor-changing neutral currents (FCNCs), in 
particular those induced by Lagrangian terms like $U_{tc}\Pi^0 \bar{t}c$. 
It has been argued previously~\cite{TC2,gustavo2} that these terms could 
be large and lead to a significant branching ratio for $\Pi^0 \to tc$. 
We will address this again in Sec.~\ref{sec:pheno}. 

The two CP-even Higgs modes in this effective 2HDM, labeled $H_{TC}$ 
and $H_{ETC}$, are known as the ``top-Higgs'' and the ``techni-Higgs'', 
respectively.  Their masses can be estimated in the Nambu--Jona--Lasinio 
(NJL) model in the large-$N_c$ approximation.  For the top-Higgs this is 
found to be on the order of $M_H \simeq 2m_t$; for the techni-Higgs it 
is much higher.  However, there is no reason to expect the NJL model to 
be correct, it only serves as a rough guide; the masses of the top- and 
techni-Higgs modes may in fact be very light.  The top-pions on the 
other hand have masses proportional to $\epsilon_t$ and the mass of the 
color octet of TC gauge bosons, $M_B$.  In the fermion bubble 
approximation this is
\begin{equation}
M^2_\Pi \, = \, {N_c \, \epsilon_t \, m^2_t M^2_B \over 8 \pi^2 f^2_\pi} \; .
\label{eq:pimass}
\end{equation}
If $M_B \sim 1$~TeV, the theory loosely predicts top-pions to lie in 
the mass range of about 100-300~GeV. Top-pions this light are disfavored 
by the data for $R_b$~\cite{R_b}, but the new physics may conspire to 
cancel the expected deviation. 

While topcolor does not give mass to the bottom quark directly, it can 
generate a contribution via instanton effects.  This contribution, 
$m^*_b \leq m_b$, is approximately, 
\begin{equation}
m^*_b \, \approx \, {3 k m_t \over 8 \pi^2} \sim \, 6.6 \, k \; {\rm GeV} \; .
\label{eq:mb}
\end{equation}
To get a limit on $k$, we use a bottom quark pole mass of $m_b \approx 4.8$ 
GeV, so that the entire $b$ quark mass would come from contribution due to 
topcolor instantons for $k \sim 0.73$.  Since Eq.~\ref{eq:mb} is only a 
rough estimate we will use $k = 0.8$ as the maximum possible value in our 
analysis.  The remaining $m_B$ contribution is assumed to come from ETC, 
via a Yukawa coupling $\epsilon_b$.  The Lagrangian terms for the ETC 
bottom Yukawa and instanton sectors are 
\begin{eqnarray}
{\cal L}_{Yuk,b} \; &
= & \; 
- \, \biggl( m^*_b + {\epsilon_b v_T \over\sqrt{2}} \biggr) \, \bar{b}b 
\nonumber \\
& & \; -{i\over v\sqrt{2}} 
       \biggl( {\sqrt{2} m^*_b\over f_\pi} v_T - \epsilon_b f_\pi \biggr)
       \; \Pi^0 \, \bar{b} \gamma^5 b \; .
\label{eq:Yuk-b}
\end{eqnarray}
For fixed $f_\pi$, this coupling depends only on $k$ ($\epsilon_b$ is 
related to $k$ by $m_b$), and has a zero at $k = 0.043$.  
Such a small non-zero value seems extraordinarily
fine-tuned so we do not consider it as a special case further.  A more
interesting special case is where ETC has flavor universal Yukawa
couplings, \ie $\epsilon_b = \epsilon_t$.  This can occur only for
very large values of the topcolor Yukawa coupling, $Y_t \gtrsim 4$
(recall for our fixed value of $f_\pi$, $Y_{t,max} \sim 4.1$).
At the lower limit of this bound, $k = 0$ and there is no topcolor
instanton-induced $b$ quark mass.


\section{Phenomenology of the model}\label{sec:pheno}

One immediately can see from Eq.~\ref{eq:Yuk-t} that the couplings of
both the top-Higgs mode and the top-pion to top quarks are enhanced by
a factor of several ($Y^{TC}_t/Y^{SM}_t \simeq 3-5$ for the top-Higgs
and $(Y^{TC}_t v_T - \epsilon_t f_\pi) / v Y^{SM}_t \simeq 3-4$ for
the top-pion) relative to the SM.  As a result, these states have a
greatly enhanced top quark loop-induced coupling to gluons. Inclusive
production, $gg\to\Pi^0$, thus occurs at a much greater rate than in
the SM.  This latter feature has been addressed previously in the
literature, briefly in Ref.~\cite{gustavo1} and in more detail in 
Ref.~\cite{gustavo2}, and we do not discuss it here.  
Furthermore, we will not discuss either the top-Higgs or the
techni-Higgs in this paper, leaving them for future
analysis~\cite{us}.  Instead, we concentrate our investigation the
neutral top-pion, which has not been examined very closely in previous
studies.

As the $\Pi^0$ is a CP-odd state, it does not couple to weak bosons at 
tree level.  This limits the production modes at a hadron collider, as 
well as the possible decay modes.  We therefore focus on single-top- 
and $t\bar{t}$-associated production, and compare the TC2 rates to 
corresponding rates in both the SM and allowed regions of the MSSM.  
We also confine our focus to the mass region $M_\Pi^0 < 2m_t$.  
For masses above the top quark pair threshold, decays to top quarks 
dominate, resulting in a rather large four top quark cross section that 
may be experimentally observable, as discussed in Ref.~\cite{spirawells}. 

All our calculations are performed with parton-level Monte Carlo using 
CTEQ4L parton distribution functions~\cite{CTEQ4} and 
$\alpha_s (M_Z) = 0.1185$.  Both the factorization and renormalization 
scales are chosen as $\mu_{f,r} = m_t + {1\over 2}M_{\Pi^0}$.  Matrix 
elements were generated with Madgraph~\cite{madgraph} by adding the TC2 
scalar sector to its model tables.  We do not consider running of the 
Yukawa couplings.  Since the values are unknown, that analysis seems 
premature.  We show only a few representative choices of the possible 
couplings, to characterize the model's general behavior.  If the $\Pi^0$ 
is observed, it will then be important to study higher order effects. 

\vspace{3mm}
\noindent{\it Decays of the neutral top-pion}
\vspace{1mm}

\begin{figure}[ht] 
\begin{center}
\includegraphics[width=7.0cm,angle=90]{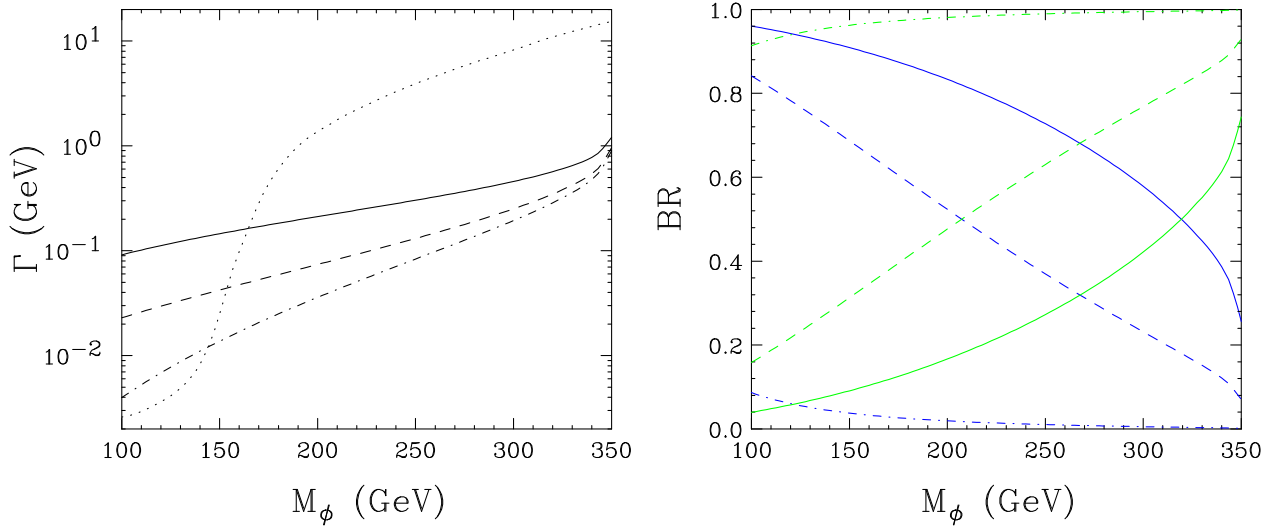}
\vspace{2mm}
\caption[]{\label{fig:br} \sl 
Total width (left) and dominant branching ratios (right) of the neutral 
top-pion, as a function of $k$ for fixed $Y_t = 4.0$. 
Shown are the curves for $k$ is 0.8 (solid), 0.4 (dashed) and 0 (dot-dashed). 
The SM Higgs total width is shown by the dotted line in the left panel. 
In the right panel, BR($b\bar{b}$) are in blue (downward sloping), and 
BR($gg$) are in green (upward sloping).  The $\Pi^0$--quark interactions 
are assumed to be flavor diagonal (see text). 
}
\end{center}
\vspace{-8mm}
\end{figure}
%

After calculating the top-pion couplings to SM particles, we evaluate
the dominant partial widths of the top-pion, decays into $b\bar{b},gg$, 
and for $M_{\Pi^0} > 2 m_t$, top quark pairs.  We assume for now that 
the $\Pi^0$--quark interactions are flavor diagonal.  The
results are shown in Fig.~\ref{fig:br} for different values of $k$ in
the $b\bar{b}\Pi^0$ instanton-induced coupling.  For $M_\phi \lesssim
150$~GeV, the top-pion has a significantly larger width than a SM
Higgs, but for $M_\phi \gtrsim 150$~GeV the top-pion is approximately
an order of magnitude narrower than a SM Higgs due to the lack of tree
level decay modes to weak bosons.  Since the top-pion width remains
less than a GeV for all masses below the top pair threshold, the
top-pion would appear experimentally as a narrow resonance, at the
limit of detector width resolution in any decay channel.  This is in
contrast to the SM Higgs, where the width already exceeds detector
resolution by about $M_H \approx 220$~GeV.  The implication is that
the total width of the top-pion can be determined only indirectly.

For $M_{\Pi^0} < 2m_t$, it is important to note that even if topcolor
does not contribute a mass to the $b$ quark (\ie $k = 0$) there is
still a small branching ratio to $b$ quarks, although for 
$M_{\Pi^0} \gtrsim 150$~GeV this quickly becomes negligible.  If instead 
$k = k_{max} \sim 0.8$, even at $M_{\Pi^0} = 100$~GeV the branching 
ratio to gluons is about $5\%$, the smallest it ever gets.  For larger
top-pion masses or more moderate values of $k$, there is typically a
rather large branching ratio to gluons.  We will later place rough
limits on what we expect $\sigma\cdot {\rm BR}$ to be for each decay
mode as a function of $Y_t$ and $k$.

For $M_{\Pi^0} > 2m_t$, the top-pion total width exceeds the SM Higgs total 
width by a factor 3-5, depending on the choice of $Y_t$. 
In this region, decays to top quark pairs dominate the width to such a degree 
that their branching ratio is effectively unity; all other decay modes may be 
ignored. 

\vspace{3mm}
\noindent{\it Single top associated production}
\vspace{1mm}

\begin{figure}[ht]
\begin{center}
\includegraphics[width=11.0cm]{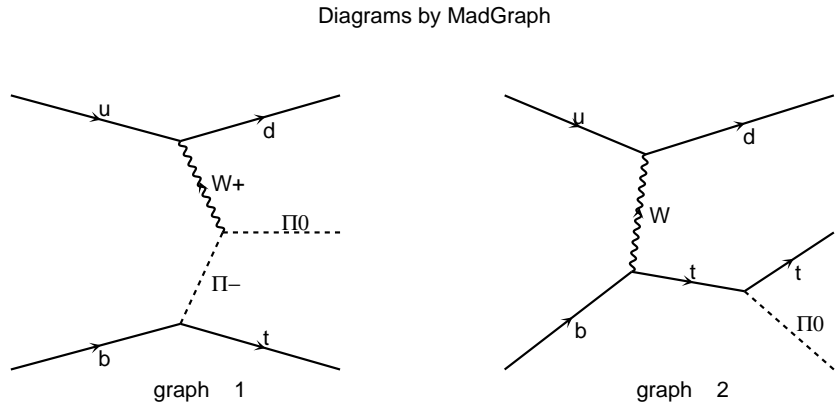}
\vspace{2mm}
\caption[]{\label{fig:singletop} \sl 
t-channel $W$ single top associated $\Pi^0$ production. 
As in the MSSM there is a strong cancellation between the two diagrams, 
leading to small, almost certainly unobservable rates.
}
\end{center}
\vspace{-6mm}
\end{figure}
%

The largest single top production cross section at the Tevatron
($\sqrt{s} = 2.0$ TeV) is s-channel production, 
$u\bar{d} \to W^* \to t\bar{b}$.  t-channel production, 
$ug \to dt\bar{b}$, dominates at the LHC ($\sqrt{s} = 14$ TeV).  
One may easily estimate that at either the
Tevatron or LHC, even if s-channel single top associated production of
a neutral top-pion is enhanced relative to the SM Higgs rate by $\sim
3^2$, an order of magnitude, this is not enough to be
observed~\cite{willen}.  t-channel production is a different story.
In this case there is a strong cancellation in the SM between the
graphs where the Higgs is radiated off the t-channel $W$ boson or off
the final state top quark, which preserves unitarity at high
energies~\cite{willen}.  Combined with the rather large background
rates, this renders SM Higgs single top associated production
unobservable at both the Tevatron and LHC.  Even in the MSSM it is
difficult to achieve a significant enough enhancement to hope for much
improved prospects.  But in TC2, the neutral top-pion cannot be
emitted from the t-channel $W$, so one would na\"{\i}vely expect
cancellations to be absent and the rate to be considerably larger.
Unfortunately, there is a $W^+\Pi^-\Pi^0$ vertex, as shown in
Fig.~\ref{fig:singletop}, which leads to a similar strong cancellation
between the diagrams, again as required by unitarity (see
Appendix~\ref{app:singletop} for details).  Since at the LHC the
top-pion production cross section is never more than a factor two
larger than for the SM Higgs, we believe this channel is not useful
and do not consider it further.

\vspace{3mm}
\noindent{\it Top quark pair associated production}
\vspace{1mm}

\begin{figure}[ht] 
\begin{center}
\includegraphics[width=6.5cm,angle=90]{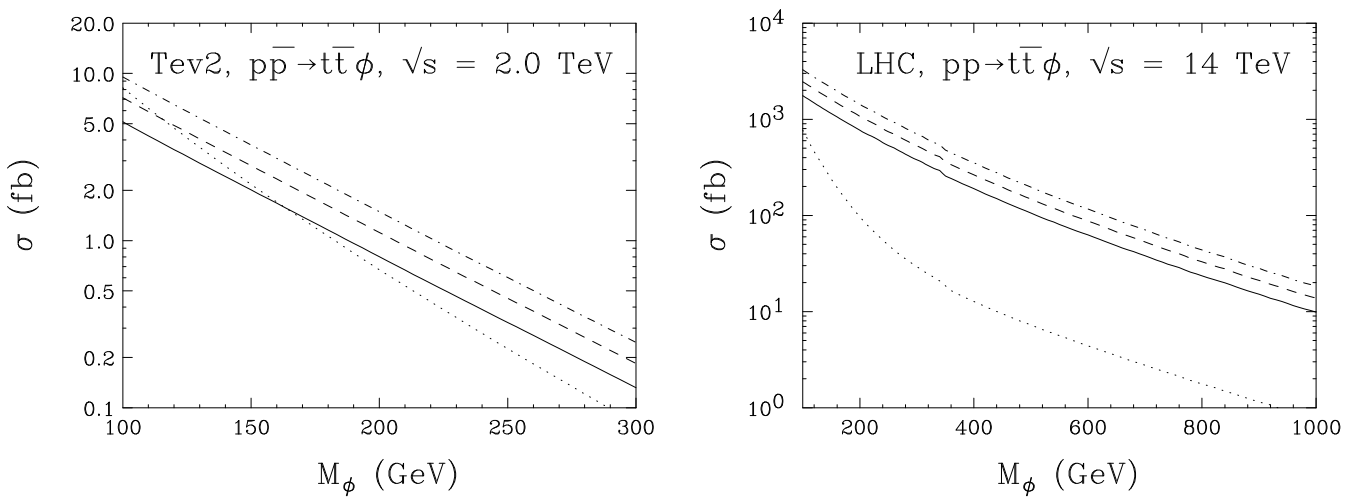}
\vspace{2mm}
\caption[]{\label{fig:xsec1} \sl 
Total $t\bar{t}\Pi^0$ v. Standard Model $t\bar{t}H$ cross sections at the 
Tevatron (left) and LHC (right). 
TC2 model input is $f_\pi = 60$~GeV and $Y_t = 3.0$ (solid), 3.5 (dashed), 
and 4.0 (dotdahsed). 
The SM cross sections are shown by the dotted curves. 
}
\end{center}
\vspace{-6mm}
\end{figure}
%

The situation is very different for $t\bar{t}\Pi^0$ production as
there are no cancellations between diagrams.  The cross section at the
Tevatron, shown in the left panel of Fig.~\ref{fig:xsec1}, is 
comparable to that for SM $t\bar{t}H$ production for 
$M_{\Pi^0} \lesssim 150$~GeV, varying within a factor of several 
smaller to few larger.  At larger $\Pi^0$ masses, 
$M_{\Pi^0} \gtrsim 150$~GeV, the TC rate is always larger, although 
the total rate is not enough to yield enough events~\cite{ttH-Tev2}. 
That the rate is only comparable rather than significantly larger, as 
one would guess from the relative magnitude of the quark-quark-scalar 
couplings, is due to a different sort of cancellation: since 
the $t\bar{t}\Pi^0$ vertex contains a $\gamma^5$, due to the CP-odd 
nature of the scalar, there is destructive interference between the
$p_{in} \cdot p_{out}$ and $m^2_t$ terms in the Dirac structure of the
amplitude.  The Tevatron runs at a partonic center of mass energy
where the terms are of comparable size, so the overall coupling
enhancement of $\approx 3^2$ is unfortunately countered; if the 
$\gamma_5$ were not present, the cross section at the Tevatron would 
be larger by an order of magnitude~\cite{us}.  

Prospects for observation of $t\bar{t}H_{SM}$ events at the Tevatron 
were initially believed to be good for 
$M_H \lesssim 135$~GeV~\cite{ttH-Tev2}, but recent NLO calculations of 
$p\bar{p}\to t\bar{t}H_{SM}$ revealed an unexpected suppression rather 
than enhancement~\cite{ttH-NLO}, which make the search much more 
difficult.~\footnote{We note as an aside that $t\bar{t}A$ production in 
the MSSM is essentially never observable, as the cross section is always 
at least one order of magnitude smaller than the $t\bar{t}H$ rate of 
equal scalar mass~\cite{spira}.} It is not yet known what the NLO result 
is for pseudoscalar production in association with top quark pairs at 
hadron colliders~\footnote{NLO results for $e^+e^- \to t\bar{t}H_{SM}$ 
turned out to be a poor guide for $p\bar{p} \to t\bar{t}H_{SM}$ at the 
Tevatron, so the known results for $e^+e^- \to t\bar{t}A_{MSSM}$ are 
also likely not so useful here.}, so we cannot make definitive comments 
on the potential observability of this channel.  The slightly lower 
cross sections for low $\Pi^0$ mass suggest that $t\bar{t}\Pi^0$ 
production is likely to be missed at the Tevatron, at least for small to 
moderate $Y_t$, but this should be viewed as a challenge to the machine 
and detector groups.  Observing or ruling out TC2 based on its neutral 
pseudoscalar content will at the very least be extremely difficult at 
the Tevatron unless the machine performs exceedingly well. 

A completely different paradigm will reign at the LHC.  From recent 
studies with detector simulation~\cite{ttH-LHC}, it is known that a SM 
Higgs of mass $M_H = 120$~GeV can be discovered in the 
$t\bar{t}H \to \ell\nu jj b\bar{b}b\bar{b}$ channel.  The studies found 
that the backgrounds can be reduced to the level of the signal, 
$S/B \sim 1/1$, yielding a statistical significance of about $12\sigma$ 
at CMS and about $10\sigma$ at ATLAS, for 100~fb$^{-1}$ of data. 
Both studies used the sample consisting of one top quark decaying 
hadronically and the other leptonically, $\approx 1/3$ of the total 
$t\bar{t}H$ event sample. 

\begin{figure}[ht] 
\begin{center}
\includegraphics[width=6.5cm,angle=90]{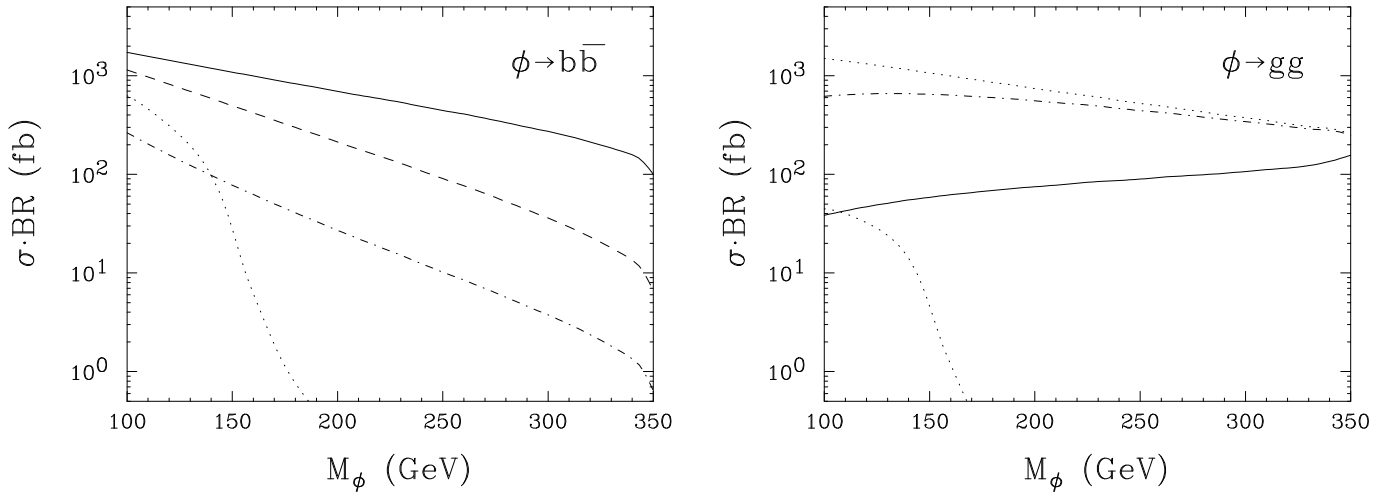}
\vspace{2mm}
\caption[]{\label{fig:xsec2} \sl 
We show the top-pion production cross sections for $Y_t = 3.0$, multiplied 
by the branching ratios to $b\bar{b}$ (left) and $gg$ (right), for various 
values of $k$: 0.8 (solid), 0.2 (dashed), 0 (dotdashed). 
The SM Higgs rates are shown by the dotted lines. 
}
\end{center}
\vspace{-6mm}
\end{figure}

We predict that any $t\bar{t}\Pi^0 ; \Pi^0 \to b\bar{b}$ rate that is 
more than half the SM rate for the same $M_\phi$ will be observable at 
greater than $5\sigma$.  Examining the left panel of 
Fig.~\ref{fig:xsec2}, for $M_{\Pi^0} = 120$~GeV this corresponds to 
$Y_t = 3.0$ and very small $k$, close to 0 (ignoring the exact zero at 
$k \approx 0.05$).  For larger $Y_t$, the top-pion signal only becomes 
stronger, as the production cross section increases faster than 
$BR(b\bar{b})$ falls off.  (This behavior holds generally for all 
$\Pi^0$ masses.)  It is manifest that any region of parameter space with 
$\sigma\cdot BR(b\bar{b}) \gtrsim 300$~fb is likewise accessible.  In 
fact the situation is much better, since the $t\bar{t}b\bar{b}$ 
background falls off very quickly with increasing $m_{b\bar{b}}$.  
However, we cannot match the level of sophistication presented in 
Ref.~\cite{ttH-LHC}, and a parton-level Monte Carlo calculation would be 
a misleading comparison, so we leave the details of reach in this channel 
to future work by detector collaborations.  We do note, however, that for 
the obviously very large region of parameter space where statistical 
significance would be $\gg 5\sigma$, the methods of Ref.~\cite{gunion} 
should also allow for confirmation of the pseudoscalar nature of the 
resonance. 

%

For larger masses $M_{\Pi^0}$, it may be possible to observe the decay 
mode $\Pi^0 \to gg$ over some region of TC2 parameter space. We know 
of no other model where this is possible.  To illustrate our claim we 
examine a few points in parameter space in Table~\ref{tbl:gg-mode}. 
Here we calculate the signal and QCD $t\bar{t}+jj$ 
backgrounds~\cite{stange} at parton level with full matrix elements, 
including the decay $\Pi^0 \to gg$.  We consider the final state where 
one top quark decays hadronically and the other decays leptonically, 
providing a hard lepton for triggering.  We do not attempt to include 
detector effects, but we do include some major detector efficiencies 
such as $b$ jet tagging ($60\%$ each) and lepton ID ($85\%$), which 
reduces the captured rates considerably.  We also apply the rather severe 
kinematic cuts needed to satisfy the experimental criteria for high 
luminosity running:
\begin{eqnarray}
p_T(j)    > 30 {\rm GeV} , & \phantom{ii} & |\eta(j)| < 4.5 \, , \nonumber\\
p_T(b)    > 30 {\rm GeV} , & \phantom{ii} & |\eta(b)| < 2.5 \, , \nonumber\\
p_T(l)    > 15 {\rm GeV} , & \phantom{ii} & |\eta(l)| < 2.5 \, , \nonumber\\
\sla{p}_T > 50 {\rm GeV} , & \phantom{ii} & \triangle R_{ij} > 0.4 \, .
\label{eq:cuts}
\end{eqnarray}
In addition, we require an additional cut $p_T > 40(50)$~GeV on the jets 
from decay of the $\Pi^0$ for $M_{\Pi^0} = 200(300)$~GeV. As the $\Pi^0$ 
is a narrow state even at the higher mass, we examine signal v. background 
in a $\pm 20$~GeV bin around the central value. 
Due to the lack of detailed detector simulation, this comparison should 
be taken only as a rough guide for the reach available in this channel. 
Our goal is to show the potential distinctive characteristics of the TC2 
model.  

\begin{table}[htb]
\begin{tabular}
{||p{2.3cm}|p{1.8cm}||p{1.5cm}|p{1.5cm}|p{1.4cm}|p{1.4cm}|p{1.4cm}|p{1.8cm}||}
$\;$ $M_{\Pi^0}$ (GeV) & $Y_t, k$ & $\sigma_S$ (fb) & $\sigma_B$ (fb) & 
$N_S$ & $N_B$ & $S/B$ & $N_S/\sqrt{N_B}$ \\
\hline
$\;$ 200 & 3.0, 0.8 &  2.2 & 680 &  140 & 62,400 & 1/440 & 0.6 \\
$\;$ 200 & 3.0, 0.2 & 16.4 & 680 & 1050 & 62,400 & 1/60  & 4.2 \\
$\;$ 200 & 4.0, 0.8 &  4.0 & 680 &  260 & 62,400 & 1/240 & 1.0 \\
$\;$ 200 & 4.0, 0.2 & 19.9 & 680 & 1280 & 62,400 & 1/50  & 5.1 \\
$\;$ 300 & 3.0, 0.8 &  3.3 & 290 &  210 & 26,600 & 1/130 & 1.3 \\
$\;$ 300 & 3.0, 0.2 & 10.6 & 290 &  680 & 26,600 & 1/40  & 4.2 \\
$\;$ 300 & 4.0, 0.8 &  9.2 & 290 &  590 & 26,600 & 1/45  & 3.6 \\
$\;$ 300 & 4.0, 0.2 & 20.6 & 290 & 1320 & 26,600 & 1/20  & 8.1 \\
\end{tabular}
\vspace{2mm}
\caption[]{\label{tbl:gg-mode} \sl 
Cross sections for the topcolor assisted technicolor signal 
$pp\to t\bar{t}\Pi^0 \to b\bar{b}l\nu jj gg$ 
(1 leptonic and 1 hadronic decay of the top quarks) and background 
$pp\to t\bar{t}jj \to \to b\bar{b}l\nu jj jj$ at the LHC, 
$\sqrt{s} = 14$~TeV. The number of events are calculated for 
300~fb$^{-1}$ of integrated luminosity, a universal efficiency factor 
of 0.31 for particle ID, and an efficiency factor of 0.7 for the mass 
bin capture of the signal only.}
\end{table}
%

Table~\ref{tbl:gg-mode} reveals that the $\Pi^0 \to gg$ decay mode 
is probably observable only for large $Y_t$ or very small $k$. 
While the number of background events is very large, $S/B$ and total 
number of signal and background events are quite similar to the SM 
$gg \to H\to\gamma\gamma$ search at the LHC, which has been shown to 
be accessible~\cite{H-LHC}.  Our estimate also makes no attempt to 
utilize the complex nature of these final states, which has elsewhere 
been shown to yield significant improvements beyond our simple 
approach~\cite{ttH-LHC}.  The Table suggests that this mode may be 
able to provide discovery coverage over regions of parameter space 
where the $\Pi^0 \to b\bar{b}$ mode is not accessible. 

If we now deviate from our assumption that the $\Pi^0$--quark 
interactions are flavor diagonal, for the $\Pi^0$ mass range 
$m_t + m_c < M_{\Pi^0} < 2m_t$ the decays $\Pi^0 \to t\bar{c},\bar{t}c$ 
can occur with substantial, even dominant branching ratio, depending 
on the magnitude of $U_{tc}$.  We are not concerned with this here, 
because in $t\bar{t}\Pi^0$ events it would lead to a spectacular 
signature of three top quarks and an additional charm jet.  There is no 
SM process that can give this, and the rate for $pp\to tttb$ at the LHC 
is less than 0.2~fb; the $b\to c$ mistagging probability would reduce 
this even further.  We will address the flavor-changing possibilities 
separately~\cite{us} and do not discuss them further here.


\section{Conclusions}

We have outlined the relevant features of topcolor assisted technicolor 
models of type I, which do not possess flavor-changing neutral currents. 
TC2 is effectively a two Higgs doublet model where one doublet is 
primarily responsible for giving mass to the top quark and the other is 
the dominant contributor to electroweak symmetry breaking.  This results 
in strongly enhanced couplings of the neutral pion mode, $\Pi^0$ 
(analogous to the CP-odd scalar $A$ in the MSSM), to top quarks.  Direct 
observation of the $\Pi^0$ via its enhanced top quark couplings would be 
confirmation that the EWSB sector realized in nature is not SM or part 
of the MSSM. 

The mass of the $\Pi^0$ is loosely expected to be fairly light, in the 
$100-300$~GeV region.  It would decay predominantly to $b\bar{b}$ or 
$gg$ final states, depending on its mass and how much instanton 
contribution to the $b$ quark mass comes from topcolor, which cannot 
be determined theoretically with much confidence and is simply 
parameterized.  

Our investigation reveals that, as in other EWSB models involving one 
or two Higgs doublets, the single top associated production mode has 
a very small, almost certainly unobservable cross section at any hadron 
collider.  $t\bar{t}$ associated production, on the other hand, benefits 
from a greatly enhanced cross section at the LHC (although not, alas, at 
the Tevatron).  We find that $t\bar{t}\Pi^0;\Pi^0 \to b\bar{b}$ events 
should be easily discernible over much of the TC2 parameter space. 
In contrast, only very small, unobservable rates for $t\bar{t}A$ 
production (as well as $t\bar{t}h$ production for $M_h \gtrsim 140$~GeV) 
are predicted for the MSSM.  Furthermore, for large $Y_t$ or very small 
$b\bar{b}\Pi^0$ coupling, and large $M_{\Pi^0}$ (but less than the top 
pair threshold), the decay $\Pi^0 \to gg$ is likely to be visible over 
the $t\bar{t}+jj$ background.  We anticipate that this will provide for 
more complete coverage of TC2 parameter space, but deserves a detailed 
detector simulation to explore fully.  For top-pion masses above the 
top pair threshold, the four top production cross section is greatly 
enhanced, although we do not address this signature. 

If TC2 is the correct model describing nature, and the top-pion is 
observed at the LHC, there is still a long way to go toward 
determining the location of the model in parameter space.  Ignoring 
the potential, there are effectively six unknowns 
($f_\pi, v_T, Y_t, \epsilon_t, k, \epsilon_b$), 
but fewer constraints: the top and bottom quark masses, EWSB $v$, 
and the $t\bar{t}\Pi^0$ production cross section times the 
branching ratio to either $b$ quarks or gluons.  Our analysis shows 
that it is not very likely for the total rate to be determinable at a 
hadron collider. But by also observing another production mode in the 
same decay channel, such as $gg\to\Pi^0 \to b\bar{b}$, one can get 
around having to know either $k$ or $\epsilon_b$.  While the number of 
unknowns is reduced to four, the number of measurements is still 
effectively three.  This leaves the system underdetermined, so that 
additional measurements would be necessary, such as the rate of 
$H_{TC}$ production times $BR(b\bar{b},gg)$ rate in either gluon fusion 
or top quark associated production.


\acknowledgements

We want to thank Chris Hill for helping us to understand topcolor assisted 
technicolor and Gustavo Burdman for useful discussions. 
Fermilab is operated by URA under DOE contract No.~DE-AC02-76CH03000.


\appendix\label{sec:app}


\section{The TC2 Lagrangian}\label{app:model}

We begin by writing the effective TC2 Lagrangian in linearized form. 
The kinetic term is
\begin{equation}
{\cal L}_{kin} \, = \, 
  \biggl( D_\mu \Phi_{TC}  \biggr)^\dagger \biggl( D^\mu \Phi_{TC}  \biggr)
+ \biggl( D_\mu \Phi_{ETC} \biggr)^\dagger \biggl( D^\mu \Phi_{ETC} \biggr),
\label{eq:L-kin}
\end{equation}
where the $SU(2)$ doublets $\Phi$ have the form 
\begin{mathletters}\label{eq:phi}
\begin{eqnarray}
\Phi_{TC}  \, & = & \, \binom{(f_\pi + H_{TC}  + i\pi^0_{TC})/\sqrt{2}}
                             {i\pi^-_{TC}},  \\
\Phi_{ETC} \, & = & \, \binom{(v_T   + H_{ETC} + i\pi^0_{ETC})/\sqrt{2}}
                             {i\pi^-_{ETC}},
\end{eqnarray}
\end{mathletters}%
and the covariant derivative is
\begin{equation}
D_\mu \, = \, \partial_\mu + i {g_Y \over 2} Y B_\mu 
   + i {g \over 2} \tau_i W^i_\mu \, .
\label{eq:Dmu}
\end{equation}
The hypercharge of the doublets is $Y = -1$, and $g$ is $g_{weak}$. 
We make the following redefinition of fields:
\begin{eqnarray}
W^\pm_\mu = {1\over\sqrt{2}}(W^1_\mu \mp iW^2_\mu), \\
W^3_\mu =   Z_\mu \cos\theta + A_\mu \sin\theta,      \\
B_\mu   = - Z_\mu \sin\theta + A_\mu \cos\theta.
\label{eq:V-fields}
\end{eqnarray}
After replacement of the physical vector boson fields, the $D_\mu \Phi_i$ term 
for each doublet will be of the form
\begin{eqnarray}
D_\mu \Phi_i \, = & 
\left(\begin{array}{c} 
{1\over\sqrt{2}}(\partial_\mu H_i + i\partial_\mu\pi^0_i) \\ 
i\partial_\mu\pi^-_i 
\end{array}\right)
& \; + \; {i g_Z\over 2} \, Z_\mu
\binom{{1\over\sqrt{2}}(v_i + H_i + i\pi^0_i)}{-i(1-2\sin^2\theta_W)\pi^-_i}
\nonumber\\
& \; + \; e A_\mu 
\left(\begin{array}{c} 
0 \\ 
\pi^-_i 
\end{array}\right)
& \; + \; {i g\over 2} 
\binom{i\sqrt{2} \, W^+_\mu\pi^-_i}{W^-_\mu (v_i + H_i + i\pi^0_i)}.
\label{eq:DmuHi}
\end{eqnarray}
where $g_Z = g/\cos\theta_W$ and $e = g\sin\theta_W$. 
After expanding the terms in Eq.~\ref{eq:L-kin}, we form orthogonal linear 
combinations of the fields $\pi^{0,\pm}_i$, 
\begin{eqnarray}
w^{0,\pm}   \; & = & \; {f_\pi \pi^{0,\pm}_{TC} + v_T   \pi^{0,\pm}_{ETC} \over v} 
\; \; \; {\rm (Goldstone \; bosons)}, \\
\Pi^{0,\pm} \; & = & \; {v_T   \pi^{0,\pm}_{TC} - f_\pi \pi^{0,\pm}_{ETC} \over v}
\; \; \; {\rm (physical \; top-pions)},
\label{eq:pions}
\end{eqnarray}
where $v^2 = f^2_\pi + v^2_T = (246 \; {\rm GeV})^2$. 

After rearrangement the Feynman rules can simply be read off.  At this
point we reverse the flow of all bosons from incoming to outgoing, to
match the treatment used in {\sc Madgraph/HELAS}.  The coefficient of
each term is the {\sc HELAS} coupling.  Table~\ref{tbl:coups1} lists
the 3-point gauge couplings for all physical fields; the Goldstone
boson and 4-point couplings are not listed for brevity.

\begin{table}[htb]
\begin{tabular}{|p{3cm}|p{5cm}||p{3cm}|p{5cm}|}
$Z^\mu Z_\mu H_{TC}$       & ${1\over 2} \, f_\pi \, g^2_Z$ &
$Z^\mu Z_\mu H_{ETC}$      & ${1\over 2} \, v_T   \, g^2_Z$ \\
$W^{+\mu} W^-_\mu H_{TC}$  & ${1\over 2} \, f_\pi \, g^2$   &
$W^{+\mu} W^-_\mu H_{ETC}$ & ${1\over 2} \, v_T   \, g^2$   \\
$Z^\mu H_{TC}  \Pi^0$ & $-{i\over 2} \, g_Z \, {v_T   \over v} \, (p^H_\mu - p^0_\mu)$ &
$Z^\mu H_{ETC} \Pi^0$ & $+{i\over 2} \, g_Z \, {f_\pi \over v} \, (p^H_\mu - p^0_\mu)$ \\
$Z^\mu \Pi^- \Pi^+$   & $g_Z \, (1-2\sin^2\theta_W) \, (p^-_\mu - p^+_\mu)$ &
$A^\mu \Pi^- \Pi^+$   & $e \, (p^-_\mu - p^+_\mu)$                       \\
$W^{-\mu} H_{TC}  \Pi^+$ & $-{i\over 2} \, g \, {v_T   \over v} \, (p^H_\mu - p^+_\mu)$ &
$W^{+\mu} \Pi^-  H_{TC}$ & $+{i\over 2} \, g \, {v_T   \over v} \, (p^-_\mu - p^H_\mu)$ \\
$W^{-\mu} H_{ETC} \Pi^+$ & $+{i\over 2} \, g \, {f_\pi \over v} \, (p^H_\mu - p^+_\mu)$ &
$W^{+\mu} \Pi^- H_{ETC}$ & $-{i\over 2} \, g \, {f_\pi \over v} \, (p^-_\mu - p^H_\mu)$ \\
$W^{-\mu} \Pi^0   \Pi^+$ & $-{1\over 2} \, g \, (p^0_\mu - p^+_\mu)$ &
$W^{+\mu} \Pi^-   \Pi^0$ & $-{1\over 2} \, g \, (p^-_\mu - p^0_\mu)$ \\
\end{tabular}
\vspace{2mm}
\caption[]{\label{tbl:coups1} 
  \sl {\sc Madgraph/HELAS} 3-point TC2 gauge couplings for the physical fields; 
      Goldstone boson and 4-point couplings are not listed.
      All bosons (charge and momentum) flow out in the {\sc HELAS} convention.}
\end{table}

Using the same scalar $SU(2)$ doublets in Eq.~\ref{eq:phi}, the Yukawa term in 
the Lagrangian is written as
\begin{equation}
{\cal L}_Y \, = \; 
- Y_t 
\biggl( \bar{\Psi}_L\Phi_{TC} \,t_R + 
        \bar{t}_R\Phi^\dagger_{TC} \Psi_L \biggr)
- \epsilon_t 
\biggl( \bar{\Psi}_L\Phi_{ETC}\,t_R + 
        \bar{t}_R\Phi^\dagger_{ETC}\Psi_L \biggr),
\label{eq:L-Yuk}
\end{equation}
where $\Psi_L$ is the $SU(2)_L$ top-bottom quark doublet as usual. 
Rearrangement of the pion fields results in the Feynman rules for the quark 
Yukawa interactions with the top-Higgs, techni-Higgs and top-pions, shown in 
Table~\ref{tbl:coups2}.

\begin{table}[htb]
\begin{tabular}{|p{3cm}|p{5cm}||p{3cm}|p{5cm}|}
$H_{TC}  \, \bar{t}_R \, t_L$ & $-{1\over\sqrt{2}} \, Y_t$ & 
$H_{TC}  \, \bar{t}_L \, t_R$ & $-{1\over\sqrt{2}} \, Y_t$ \\
$H_{ETC} \, \bar{t}_R \, t_L$ & $-{1\over\sqrt{2}} \, \epsilon_t$ & 
$H_{ETC} \, \bar{t}_L \, t_R$ & $-{1\over\sqrt{2}} \, \epsilon_t$ \\
$\Pi^0 \, \bar{t}_R \, t_L$ & $+{i\over v\sqrt{2}} \, (Y_t v_T - \epsilon_t f_\pi)$ & 
$\Pi^0 \, \bar{t}_L \, t_R$ & $-{i\over v\sqrt{2}} \, (Y_t v_T - \epsilon_t f_\pi)$ \\
$\Pi^- \, \bar{t}_R \, b_L$ & $+{i\over v} \, (Y_t v_T - \epsilon_t f_\pi)$ & 
$\Pi^+ \, \bar{b}_L \, t_R$ & $-{i\over v} \, (Y_t v_T - \epsilon_t f_\pi)$ \\
\end{tabular}
\vspace{2mm}
\caption[]{\label{tbl:coups2} 
  \sl {\sc Madgraph/HELAS} Yukawa quark-quark-scalar TC2 couplings. 
      $Y_t$ is the large topcolor top quark Yukawa, and $\epsilon_t$ is the 
      ETC Yukawa giving a small contribution to the top quark mass. 
      All bosons (charge and momentum) flow out in the {\sc HELAS} convention.}
\end{table}
%


\section{Single-top associated $\Pi^0$ production}\label{app:singletop}

To examine the analytical behavior of single top associated $\Pi^0$ production 
at hadron colliders we write the amplitudes for the two Feynman graphs in 
Fig.~\ref{fig:singletop} in the effective-$W$ approximation as in 
Refs.~\cite{willen,bordes}: 
\begin{eqnarray}
\bar{u}_t \, \biggl(  {i\over 2v} \, 
                      (Y_t \, v_T - \epsilon_t f_\pi) (1-\gamma_5) \biggr) \, u_b \,
{-1\over (p_t - p_b)^2 - M^2_\Pi} \, {g\over 2} \, 
(p^+_\mu - p^0_\mu) \, \epsilon^\mu, \\ 
\nonumber \\
\bar{u}_t \, \biggl( -{i\over v\sqrt{2}} \, 
                      (Y_t \, v_T - \epsilon_t f_\pi) \, \gamma_5 \biggr) \,
{-(\sla{p}_b + \sla{k} + m_t)\over (p_b + k)^2 - m^2_t} \, {g\over 2\sqrt{2}} \
\gamma_\mu (1-\gamma_5) \, u_b \, \epsilon^\mu .
\label{eq:amps}
\end{eqnarray}
The same couplings appear in both diagrams.  Using the high energy
limit $\epsilon^\mu = k^\mu / M_W + {\mathcal O}(M_W/k^0)$, the first
term reduces completely to the couplings coefficient and a simple
Dirac structure,
\begin{equation}
- \, {ig\over 4vM_W} \, (Y_t \, v_T - \epsilon_t f_\pi) \; 
\bar{u}_t (1-\gamma_5) u_b \; 
= \; - \, C \, \bar{u}_t (1-\gamma_5) u_b \; .
\label{eq:term1reduc}
\end{equation}
The second term reduces almost as neatly, in the quite reasonable 
approximation for high energy scattering that $m_b \sim 0$: 
\begin{equation}
C \, \biggl[ \, 
\bar{u}_t \, (1-\gamma_5) \, u_b 
\; - \; 
{m_t \over (p_b + k)^2 - m^2_t} \, 
\bar{u}_t \, \sla{p}_\Pi \, (1-\gamma_5) u_b 
\, \biggr] \; .
\label{eq:term2reduc}
\end{equation}
The first term of Eq.~\ref{eq:term2reduc} cancels the contribution from 
the first graph in Eq.~\ref{eq:term1reduc}, leaving a term that 
satisfies the unitarity constraint at high energy~\cite{willen,bordes}.


\bibliographystyle{plain}

\end{document}